# Electrospun nanodiamond-silk fibroin membranes: a multifunctional platform for biosensing and wound healing applications

Asma Khalid[1,2*], Dongbi Bai[1], Amanda N. Abraham[1,2], Amit Jadhav[3], Denver Linklater[1], Alex Matusica[4], Duy Nguyen[1], Billy James Murdoch[1], Nadia Zakhartchouk[1], Chaitali Dekiwadia[1], Philipp Reineck[1,5], David Simpson[6], Achini K. Vidanapathirana[2,5,7], Shadi Houshyar[8], Christina A. Bursill[2,5,7], Elena Ivanova[1], and Brant Gibson[1,2].

1. School of Science, RMIT University, Melbourne, VIC, Australia
2. Australian Research Council (ARC) Centre of Excellence for Nanoscale BioPhotonics (CNBP)
3. School of Fashion and Design, RMIT University, Melbourne, VIC, Australia
4. School of Computer Science, Engineering and Mathematics, Flinders University, Clovelly Park SA, Australia
5. Vascular Research centre, Lifelong Health, South Australian Health and Medical Research Institute (SAHMRI), Adelaide, SA, Australia
6. School of Physics, University of Melbourne, Melbourne, VIC, Australia
7. Adelaide Medical School, Faculty of Health and Medical Sciences, University of Adelaide, Adelaide, SA, Australia
8. School of Engineering, RMIT University, Melbourne, VIC, Australia

*asma.khalid@rmit.edu.au

**Abstract:** Next generation wound care technology capable of diagnosing wound parameters, promoting healthy cell growth and reducing pathogenic infections noninvasively would provide patients with an improved standard of care and an accelerated wound repair mechanism. Temperature is one of the indicating biomarkers specific to chronic wounds. This work reports a hybrid, multifunctional optical material platform - nanodiamond-silk membranes as bioinspired dressings capable of temperature sensing and wound healing. The hybrid structure was fabricated through electrospinning and formed 3D sub-micron fibrous membranes with high porosity. The silk fibres are capable of compensating for the lack of extracellular matrix at the wound site, supporting the wound healing process. The negatively charged nitrogen vacancy ($NV^-$) color centres in nanodiamonds (NDs) exhibit optically detected magnetic resonance (ODMR) properties and act as fluorescent nanoscale thermometers, capable of sensing temperature variations associated to the presence of infection or inflammation in a wound, without physically removing the dressing. Our results show that the presence of NDs in the hybrid ND-silk membranes improve the thermal stability of silk fibres. The $NV^-$ color centres in NDs embedded in silk fibres exhibit well-retained fluorescent and ODMR properties. Using the $NV^-$ centres as fluorescent nanoscale thermometers, we achieved temperature sensing at a range of temperatures, including the biologically relevant temperature window, on cell-cultured ND-silk membranes. An enhancement in the temperature sensitivity of the $NV^-$ centres was observed for the hybrid materials. The hybrid membranes were further tested *in vivo* in a murine wound healing model and demonstrated biocompatibility and equivalent wound closure rates as the control wounds. Additionally, the hybrid ND-silk membranes showed selective antifouling and biocidal propensity toward Gram-negative *Pseudomonas aeruginosa* and *Escherichia coli*, while no effect was observed on Gram-positive *Staphylococcus aureus*.

## 1. Introduction

A biodegradable dressing capable of both wound healing and monitoring represents a next level multifunctional platform for improved current wound care technologies. Electrospinning of polymers generates a 2D or 3D structure of nano and sub-micron sized fibres, in the form of membranes, with high porosity and high surface area, mimicking the extracellular matrix needed for healing. These fibrous membranes are highly suitable as wound dressings to protect the wound from bacteria while keeping the flow of oxygen and nutrients through them [1-2].

However, the human body can be extremely selective regarding implants, and in many cases will quickly reject or react poorly to most polymers used as electrospun membranes [1]. Even the highly biocompatible polymers are not always suitable. Some dressings require constant maintenance, while others need to be replaced every few days which is labour intensive, may disrupt healing or may cause pain and discomfort [3]. Regenerated silk protein solution[4-5] obtained from *Bombyx mori* silk cocoons possesses good biocompatibility, excellent optical properties and flexibility to functionalize with a wide variety of optically active nanoparticles, therapeutic agents and biochemical markers[6-8]. The optically transparent silk solution can be transformed into fibres, films, gels and scaffolds [9] that are elastic [10], biodegrade and are easily metabolized in the body[11]. This regenerated silk fibroin can also be electrospun into biocompatible fibrous membranes and scaffolds for implantable devices[12].

While it is important to keep the wound protected during healing, the diagnosis of any infection and inflammation are equally crucial. Traditional methods involve a check for the presence of redness, heat detected by palpation and swelling [13]. These visual signs appear when inflammation and infection have advanced too far making therapies or interventions substantially more challenging.

Realizing a new technology that can aid clinicians to detect infections non-invasively would be beneficial and cost effective without the painful procedure of dressing removal. A chronically infected wound has a specific elevated thermal gradient range of +4 ºC to 5 ºC, compared to the normal tissue [13-14]. Nanodiamonds (NDs) with photostable negatively charged nitrogen vacancy (NV⁻) centres, shown in Figure 1(a), provide a possible avenue towards highly localised, non-invasive optical detection of temperature changes [15-17]. The electron spin energy levels of the NV⁻ centre in diamond are influenced by temperature T, as shown in Figure 1(a) and (b). This enables the zero-field splitting parameter D as a function of temperature: $D(T) = D_0 - \alpha T$, with $D_0 \sim 2.87$ GHz at room temperature and $\alpha = 74$ kHz/°C [4]. The D parameter can be initialized and readout via optically detected magnetic resonance (ODMR) technique with relative ease-of-operation [5,7], leading to NV-based thermometry in a temperature changing range of -150°C up to 400 °C [17]. Hence the NV⁻ centre can measure the local thermal environment at a spatial resolution down to sub-micron scale making it an attractive tool for high resolution spatial thermometry in infection monitoring applications [18-21]. Moreover, the NV⁻ centres record the temperature measurement of the area in thermal equilibrium with the ND[19] which minimizes the effects of environmental changes[19].

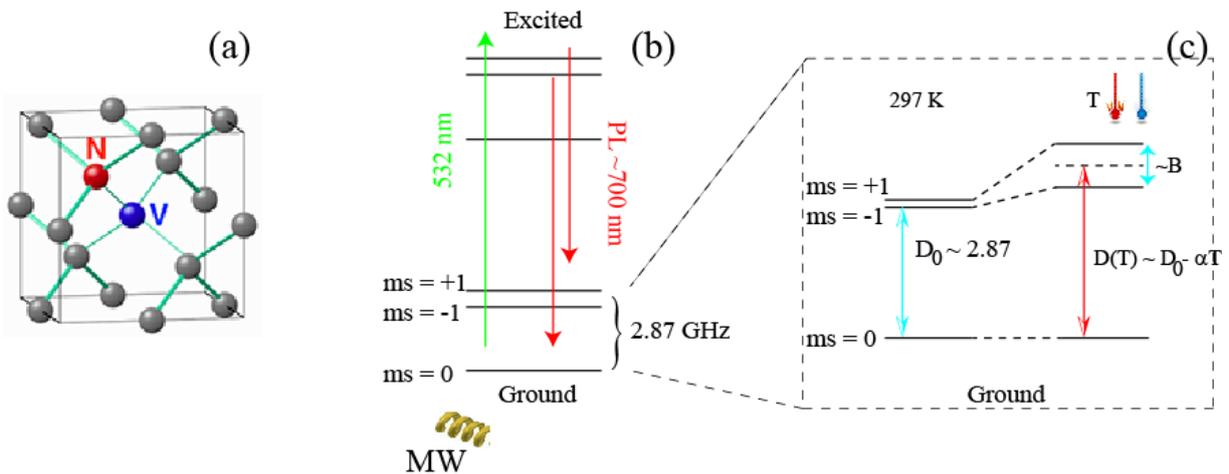

**Figure 1:** (a) Structure of NV color centre in a diamond crystal lattice. (b) Simplified energy levels of the excited and ground states of the NV⁻ centre, showing optical and MW transitions. (c) Energy diagram of the NV centre's ground spin level as a function of temperature. Electron spin energy levels are influenced by temperature T. The zero-field splitting, $D_0 \sim 2.87$ GHz for single crystal diamond at room temperature varies depending on temperature; $D(T) = D_0 - \alpha T$, where $\alpha = 74$ kHz/°C [15].

Here we present the combination of two unique optical materials: NDs and silk, electrospun into membranes as a multifunctional platform for biosensing and wound healing applications.

Electrospinning of silk generated sub-micron thick fibres with high porosity essential for breathable wound dressings. The fluorescent and spin properties of the NV$^-$ centre in NDs allow high-precision microscopic thermometry for wound sensing. Silk enables the attachment and growth of healthy cells inside the scaffold, while promoting wound healing.

In Section 2.1, we introduce the fabrication of hybrid ND-silk membranes. All samples produced for this work contained ~85% silk-fibroin and 15% polyethylene oxide (PEO) polymer by weight. Since the PEO concentration is the same for all the samples, we will refer to them as silk only (without NDs) and ND-silk (with NDs) membranes instead of silk-PEO blend membranes in the rest of the manuscript. Moreover, regenerated silk fibroin is stated as silk in the rest of the manuscript. We have discussed the structural and fluorescence properties as well as thermal stability of ND-silk membranes relative to the silk only membranes (without NDs) in Sections 2.2 and 2.3. We demonstrate the application of ND-silk hybrid membranes for temperature sensing in an *in vitro* assay in Section 2.4. The *in vivo* compatibility of the ND-silk hybrid membranes is presented in a murine model of wound healing in Section 2.5. Finally, the antibacterial resistance and bactericidal efficiency of the membranes against common pathogenic bacteria are reported in Section 2.6. Our results indicate a cost-effective combination of materials that can aid clinicians to detect infections remotely without the painful procedure of dressing removal.

## 2. Results and discussion

### 2.1 Electrospun hybrid fibres

Regenerated silk fibroin was extracted from *Bombyx mori* silk cocoons, and the protein purification was performed using the molecular cut off method as reported in literature [6, 9], yielding a pure silk fibroin aqueous solution with a concentration of 65 mg/mL. A higher viscosity and concentration of the polymer is generally required for electrospinning in order for the polymer molecules to entangle and form a jet. Hence, small volume of polyethylene oxide (PEO) in water was mixed with silk aqueous solution for this protocol, to promote fibre formation by increasing the overall polymer concentration. As also reported in earlier studies[22], the addition of PEO as a backbone carrier polymer reduces the repulsion of the biopolymer chains due to the formation of hydrogen bonds between the -OH bond of the PEO and the water molecules. This ultimately reduces the polyelectric effect in biopolymers (like silk), allowing for continuous and stable fiber production[22]. A solution of 50 mg/mL PEO in water was added to the as prepared silk solution (65 mg/mL) to generate an aqueous solution of 87% silk and 13% PEO by weight in the final mixture.

To compare the properties of ND-silk hybrid membranes with the silk only samples, we produced two different mixtures by PEO to (i) silk only and with (ii) 0.5 mg/mL NDs and silk. The schematic of the setup of the electrospinning process is illustrated in Figure 1S of the supporting information. The fibres were collected until a desired thickness of ~ 300 μm was achieved. The membranes deposited on glass cover slips were then left in a 90 v/v% methanol/water solvent for 10 min to induce crystallization of β-sheets, resulting in water insoluble membranes. The membranes were finally washed and dried for following characterizations.

**X-ray photon spectroscopy**

An X-ray photon spectroscopy (XPS) analysis of the silk and ND-silk membranes was performed to estimate the surface composition of the membranes before and after methanol treatment step. The XPS spectra with C, N, and O peaks is shown in Figure S2(a) of the supplementary information. The chemical assignment for the resultant high resolution $C_{1s}$ peaks for silk, ND-silk before and after methanol treatment are shown in Figure S2(b-e). The peak fitting results for the Silk (with PEO) & ND-Silk (with PEO) samples before/after methanol treatment are displayed in Table 1.

The peak positions in the XPS data (Figure S2) are characterized based on the binding energies of C–C(sp$^2$)/C–H (284.8 eV), C–N (285.9 eV) and C=O (287.9 eV) from the literature [23]. Compared with the carbon peaks of the silk and ND-silk membranes, the chemical bonds of the methanol treated membranes varied in peak intensities, as clearly visible in Figure S2 and Table 1. The bonds of C=O, C-C(sp$^2$) and C-H strengthened after methanol treatment, while the C-N bond was reduced noticeably. The C-N bonds mainly exist in the protein of the silk fibres and the reduction is likely caused by the lack of solubility of silk fibroin in methanol. As a result, it is expected that the PEO migrates to the surface of the membranes, increasing the C-C, C-H and C=O content on the surface.

The XPS spectroscopy revealed that both silk and ND-silk membranes demonstrated similar surface composition. The relative increase in sp$^2$/C-H bonding at the surface after the treatment indicates some extra methanol residues. Other than that, no additional impurities were found, and the methanol treatment contributes to enhancing PEO content on the surface.

Table 1: Peak fitting results for the silk & ND-silk blend membranes before and after methanol treatment. The C-H peak is close enough to the sp$^2$ peak to be considered indistinguishable.

| Sample | Peak | Position (eV) | Conc. (%) |
|---|---|---|---|
| **ND-Silk (before)** | C-N | 286.25 | 38.23 |
| | C=O | 287.87 | 23.2 |
| | C-C (*sp$^3$*) | 285.72 | 15.06 |
| | C-C (*sp$^2$*) / C-H | 284.7 | 23.51 |
| **ND-Silk (after)** | C-N | 286.14 | 24.15 |
| | C=O | 287.88 | 28.26 |
| | C-C (*sp$^3$*) | 285.7 | 18.1 |
| | C-C (*sp$^2$*) / C-H | 284.67 | 29.49 |
| **Silk (before)** | C-N | 286.16 | 35.87 |
| | C=O | 287.81 | 22.83 |
| | C-C (*sp$^3$*) | 285.75 | 16.93 |
| | C-C (*sp$^2$*) / C-H | 284.7 | 24.37 |
| **Silk (after)** | C-N | 286.19 | 27.9 |
| | C=O | 287.96 | 26.46 |
| | C-C (*sp$^3$*) | 285.73 | 15.15 |
| | C-C (*sp$^2$*) / C-H | 284.78 | 30.5 |

## 2.2 Structural and fluorescence analysis

The morphology of the electrospun membranes as well as their fluorescence properties, were examined using scanning electron microscopy (SEM) and confocal fluorescence microscopy as demonstrated in Figure 2. The figure compares the structural and fluorescence properties of ND-silk with silk only membranes. As demonstrated by the quantitative analysis of the SEM images of Figure 2(a), (c) via MATLAB, both samples consisted of randomly oriented fibres with diameters in the range of 500-800 nm, making highly porous membranes with a pore size of 0.8-1 μm. The inter-fibre spacing or porosity was measured to vary from 230-800 nm.

The addition of NDs had no noticeable difference in the fibre dimeter or membrane pore size, as indicated by SEM images. However, the confocal microscopic images of Figure 2(b), (d) showed significant difference in the fluorescence characteristics for the ND-silk and silk membranes.

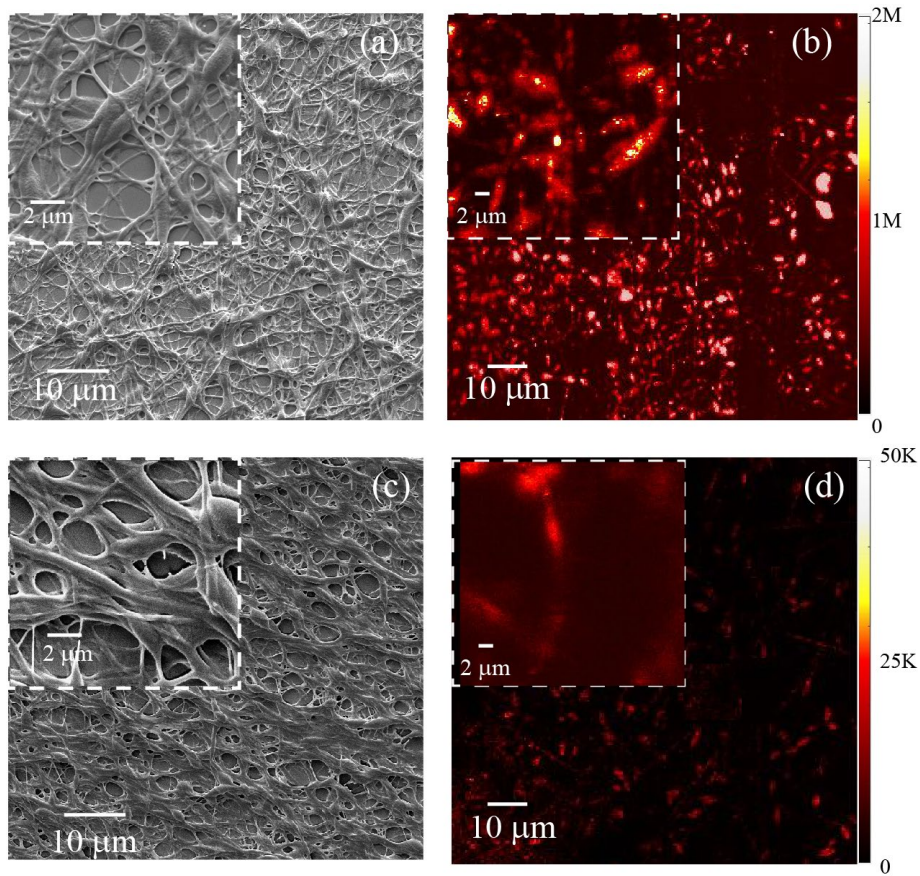

**Figure 2:** (a) SEM and (b) 50×50 μm² confocal fluorescence map of post-methanol treated ND-silk membrane deposited on glass cover slip. (c) SEM and (d) confocal fluorescence for silk only membranes. The SEM and confocal images show different areas of the same sample. The insets inside the dashed white boxes shows higher magnification scans of the samples both for SEM and confocal images.

Upon excitation with 532 nm and a pump power of only 10 μW, the ND-silk sample showed bright fluorescent from the embedded NV⁻ emitters typically show photocounts of 500k-2 M counts/s. The inset of the figure (inside the dotted white box) shows a zoomed in area with bright red and yellow fluorescence from the NDs. In contrast, the autofluorescence of the silk fibres are three order of magnitude weaker (20-50 k counts/s) under the same laser illumination power, as shown in the confocal map of Figure 2(d). The inset of the Figure 2(d) inside the dotted white box shows significantly low silk autofluorescence in the absence of NDs. Hence the NDs can brightly fluoresce inside the silk fibres and can be optically imaged with low excitation power.

**2.3 Optical characterization and temperature sensing**

We performed temperature sensing with the hybrid ND-silk materials in an *in vitro* assay. Human skin keratinocytes (HaCaT) were grown on the surface of the ND-silk membranes (on glass coverslips) for 48 hours. The growth of healthy cells on the surface of the ND-silk membranes indicates the *in vitro* compatibility and potential of the membranes for cell regeneration, as indicated in Figure 3(a). (Details on cell viability are provided in supplementary information, Figure S3). After 48 hour growth, the cells were fixed, and the hybrid membranes, as cell scaffolds, were examined via a custom-built scanning fluorescence confocal microscope. A schematic of the confocal setup is shown in Figure 3(b). We used a continuous-wave 532 nm laser as the excitation light. The pump power was measured to be 10 μW. As a result, a 100×100 μm² fluorescence map of the cell cultured ND-silk is shown in Figure 3(c), which is collected through a 532 nm dichroic mirror and a 532 nm notch filter to separate the NV's fluorescence from the excitation laser. The fluorescence map of the figure shows a fixed keratinocyte cell growing in the pores of the membrane incorporated with emitting ND spots. The typical photoluminescence spectrum from the ND emitters is shown in Figure 3(c). A characteristic NV⁻ spectrum was obtained with the zero-phonon line (ZPL) at 638 nm and a broad phonon sideband emission with a maximum around 690 nm. The neutral NV$^0$ centres with ZPL at ~576 nm was also measurable from our optical analysis.

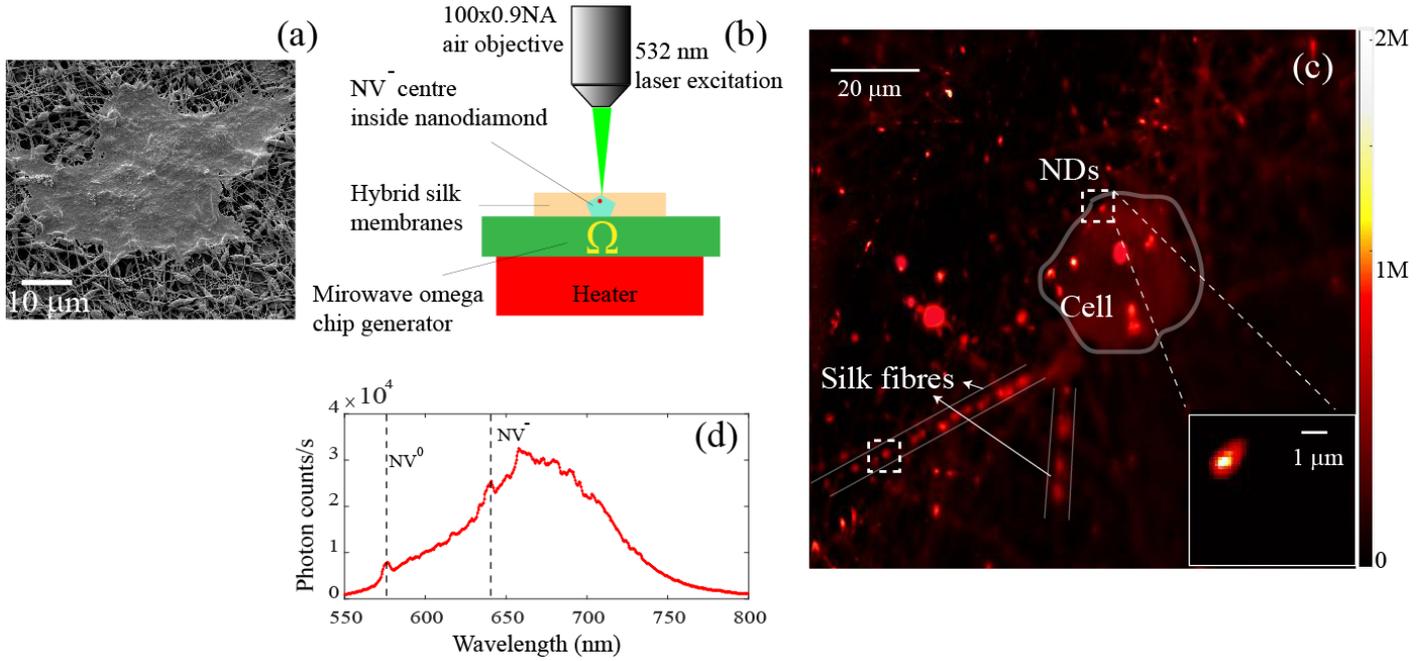

**Figure 3:** (a) Scanning electron microscope images of HaCaT cells grown on ND-silk fibrous membranes after 48 h cell culture. (b) Schematic of the setup used for optical excitation of NV$^-$ centres and temperature detection in NDs, embedded in silk. MW field is introduced with an omega chip resonator and the temperature is varied with a thermoelectric Peltier heater. (c) A 100×100 µm$^2$ fluorescence map of the cell cultured ND-silk membranes on glass cover slip. The central rounded region is one of the fixed cell growing on the hybrid membrane. The solid straight lines represent the silk fibres with embedding NDs. The dashed boxes show two representative NDs and the inset inside the solid outlined box presents a magnified 10×10 µm$^2$ scan of the selected area of interest. (d) The fluorescence spectrum obtained from emitting NDs, confirming the presence of NV centres.

The temperature sensing capability of NDs inside the cell-cultured silk-ND membranes was examined by subjecting the emitting NDs to scanning microwave (MW) frequency while monitoring the fluorescence intensity change. MW radiation for spin manipulation was applied by using an omega-shape resonator. A Peltier stage was used to test an extended range of temperatures from 20-50 °C that includes the biologically relevant range of 36-42°C[24]. The wider temperature range was applied to examine and verify the larger dD/dT shift for NV temperature sensing in the hybrid ND-silk material. A portable thermocouple was clamped with the membrane to track the temperature variation with a resolution of 0.4 °C. The measurement results are shown in Figure 4(a) where the blue data shows the ODMR spectrum detected from the ensemble NV$^-$ centres inside the silk fibres at room temperature (22 °C). The measured ODMR data was fitted by a single Lorentzian. A clear dip at ~2871 MHz is observed for the fluorescence in $m_s = 0$ to $m_s = -1$ and $+1$ transitions of NV$^-$ centre in diamond, which reveals the value of zero field splitting parameter $D$ at room temperature [16]. With increase in temperature, the zero field splitting energy shifts towards lower frequencies, bringing the dip in the ODMR spectrum to lower frequencies. The red data in Figure 4(a) indicates a lower $D$ value of ~2866 MHz at 55 °C. The dip shift in the NV$^-$ fluorescence thus enables a convenient optical readout for local temperature changes.

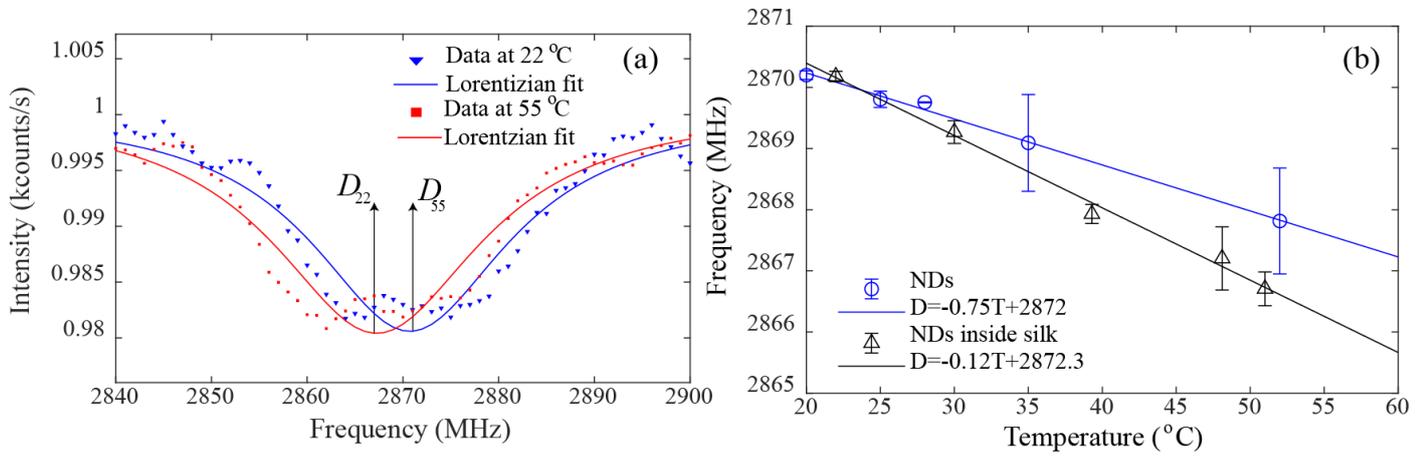

**Figure 4:** (a) ODMR spectra recorded from NV⁻ centres embedded in cell-cultured ND-silk membranes. Fluorescence emission intensity from NV centres were measured as a function of MW frequency at an ambient temperature of 22°C (blue) and increased temperature of 55 °C (red). The dot points are the experimentally obtained data, the solid curves represent the Lorentzian fit to the data of the corresponding colour. The zero filed splitting parameter *D* in the ODMR spectrum shifts with temperature. Black arrow points out the fluorescence dip at 22 and 55 ºC, respectively. (b) The *D* values plotted as a function of temperature for ND particles on their own (blue) and ND particles inside silk membranes (black). Circles and triangles show the experimental data and the solid line corresponds to the linear fit. At each temperature point, 3-5 readings were recorded and the average central frequency and the error in the mean value is plotted with red triangles.

We carried out temperature sensing in the window of 22 °C to 55 °C by using different NV⁻ emitting spots inside silk fibres. Black triangles and their linear fit in Figure 4(b) show the calibration for one typical NV emitting spot. The resonant frequencies extracted from the series of ODMR spectra obtained at elevated temperatures formed a linear function, offering a reliable calibration for temperature measurements. For this representative NV emitting spot embedded in cell-cultured silk membrane, the frequency shift per degree rise in temperature (*dD/dT*) revealed a value of −120±15 kHz/°C. We further examined different NV emitting spots in the ND-silk membranes and recorded the *dD/dT* parameter in the value between 87 kHz/°C and 162 kHz/°C, with an average of 118±25 kHz/°C. Additional temperature sensing data for NDs embedded in cell-cultured silk fibres is presented in Figure S4 in supplementary information. We point out that this parameter is larger than the typical temperature dependence of zero field spitting *D* for NV⁻ centres [15]. As a comparative study, we characterized the frequency response of hothe NDs only sample by directly testing the 100 nm diamond particles without silk coating. The calibration data revealed a *d*D/*dT* values of −75±3 kHz/°C, which agrees with the typical frequency shift of NV centres in diamond as reported in literature [15-17]. Blue circles and their linear fit in Figure 4(b) illustrate the temperature sensing data from ND-only sample.

The higher dD/dT value presented of NV⁻ centre in hybrid membrane is an added benefit as it enables more pronounced frequency shift in the detected fluorescence spectrum, making temperature testing and tracking easier for biological measurements. Our hypothesis on the enhanced dD/dT value is that by changing the surrounding temperature, silk will spatially interfere with the diamond lattice of the embedded NDs, increasing the surface pressure. This surface pressure, induced by movement of silk fibres during heating, possibly affects the spin properties of nitrogen-vacancy centres in diamond[25], and therefore induces larger than average frequency change in the detected ODMR spectra. However future experimental and theoretical work is required to reveal the fundamental physics behind this experimentally observed phenomenon.

We determined the temperature sensitivity in our hybrid ND-silk membrane by $\eta \approx \Delta \nu / C \sqrt{I_{PL}} \cdot (dD/dT)^{-1}$ [21], where $\Delta \nu$ is the linewidth of ODMR transition, *C* is the optical contrast of ODMR dip, and $I_{PL}$ is the NV⁻ fluorescence emission rate, *d*D/*dT* is the temperature dependence of zero field splitting. As a result, the experimental data shown in Figure 4 demonstrated a temperature sensitivity of ~1 K/√Hz for localized NV⁻spin characterization for the ND-silk hybrid membranes. The temperature sensitivity can be further enhanced by applying pulsed laser excitation during sensing process [16, 26]. Larger size diamonds particles [27] with increased NV$^-$ centre density [28] can be studied for higher sensitivity and brighter temperature sensing. Additionally, we have demonstrated that widefield ODMR spectroscopy and temperature sensing is also feasible with the ND-silk membranes (Figure S6 in supplementary information), which paves the way for new-generation, non-invasive temperature sensing in cell environment without the limitation of confocal settings. Our optical characterization also confirmed that the excitation laser does not induce additional heating effects for the NV$^-$ centres in ND-silk membranes (supplementary information, Figure S5).

## 2.4 Thermal stability

Prior to wound testing, the thermal properties of the hybrid ND-silk hybrid membranes were investigated to examine their thermal stability. This was performed using differential scanning calorimetry (DSC) as presented in Figure 5. DSC is a technique used to study the thermal transition of a polymer or hybrid polymer scaffolds on heating [29].

We performed the DSC analysis to check the thermal stability of silk and ND-silk membranes. The silk sample (red) was thermally stable up to 45 ºC while ND-silk hybrid (dotted blue) was stable in the whole biologically relevant window of 20-50 °C as shown in the first heating cycle of Figure 5(a).

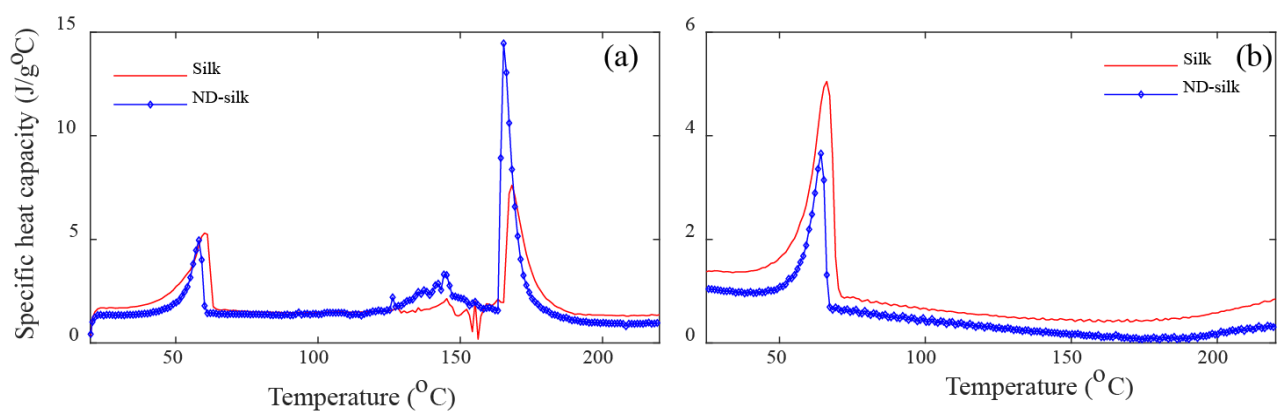

**Figure 5:** DSC curves for silk and SF-ND electrospun membranes. The specific heat capacity as a function of temperature during the (a) first and (b) second heating cycles for the same set of samples.

Hence the presence of NDs added to the stability of the membranes. The first endothermic peak was observed between 45-60 °C for the silk and 50-60 °C for the ND-silk membranes, which is representative of the melting of PEO in the samples [30]. A low-intensity DSC signal from 110-160 °C is attributed to the moisture content trapped in the regenerated fibroin [31]. The water content is variable for these materials, and there are two types of water exist in the sample; intermolecular water and free water [32-33]. The peaks below 110 °C is mostly related to the water loss, of free water present in the samples, and the sharp spike peaks in the range 130°C to 160 °C represent the water loss of intermolecular/bound water to silk. To be able to identify the water removal behaviour from silk materials, the sealed pan was not used in this study. Therefore, the evaporated water could escape and results in a series of small sharp spike peaks in the curve, instead of broad peak around this area, 140 - 155 °C. The second main and sharp endothermic peak was appeared at the glass transition temperature of silk, which involves the denaturation of the silk and resulted in the β-sheets formation [31-33]. The glass transition of the silk is water-dependent and can change based on the amount of water that exists in the sample. The water in the silk can act as the plasticizer and promote β-sheets crystallization via helix-helix interaction. This endothermic peak is just above 150 °C for silk only sample in compared with SF-ND at ~ 177 °C, which has moved to a higher temperature, due to the presence of ND and PEO in the sample [32-33]. Another reason would be due to the hydroscopic nature of PEO and ND, which attracts the water away from the silk and results in higher denaturation temperature since water assists denaturation process.

Furthermore, the data for this second peak also illustrates that significantly more energy required for denaturisation of silk fibroin in ND-silk (blue) in comparison with silk alone (red), due to presence of NDs. Hence NDs, due to their high thermal conductivity and nucleation effect, contribute to better thermal stability of the hybrid membranes in the biologically relevant window as well as at higher temperatures. The denaturation of silk fibroin polymer (between 150-180 °C) is irreversible, hence no endothermic peaks appeared in second heating cycle for both silk and ND-silk, as indicated in Figure 6(b).

## 2.5 Wound closure and healing

Silk and ND-silk hybrid membranes were tested in an established murine model of wound healing[34] for ten days. No significant differences in weight, food consumption or behaviour were noted between mice with silk or ND-silk membranes placed on their wounds. We observed that the silk membranes degraded faster (~6 days post wound surgery) than the ND-silk (~8 days post-wound surgery), which is consistent with the findings of the previous section where NDs are found to provide reinforcement to the silk fibres. There were no differences in wound healing rates in the early stages of wound healing up to day 6 post-wounding (Figure 6 (a,b). The wounds with silk healed relatively faster at day 7 (18.0 ± 9.51%) and day 8 (14.75± 10.02 %), when compared to phosphate buffer saline (PBS) treated control wounds as shown in Figure 6(a). When considering the area under curve (AUC, inset Figure 6 (a), this overall positive effect on the rate of wound healing with silk is consistent with previous studies reporting compatible, improved outcome of silk-based wound dressings [35-36]. However, there were no significant differences in wound closure between the ND-silk membrane treated wounds and their PBS controls at early, late or for the overall stages of wound healing, as shown in Figure 6(b). This difference and particularly the absence of the positive wound healing effect of silk in the ND-silk group can be attributed to earlier degradation of silk compared to ND-silk, biological variability of the two groups studied and the possibility that NDs mount a minor inflammatory response. It is important to note that the presence of NDs does not negatively affect the wound healing process, compared to the PBS control.

Laser Doppler imaging was used to determine wound blood flow perfusion, a marker of wound angiogenesis that is essential for healing. Overall, no differences were observed in wound blood flow perfusion between silk or ND-silk membrane treated wounds, when compared to their respective PBS controls as indicated by Figure 6(c) and (d), further supporting the suitability of ND-silk for sensing applications in wound dressings.

The wound cellular infiltrate in the granulation site of the wounds was measured using the histological sections represented in Figure 6 (e-g) and reflects inflammatory-driven recruitment of cells to the wounds. There were no significant differences in the cellular infiltrate between wounds with silk membranes and PBS control wounds as demonstrated in Figure 6(h). However, a non-significant increase in the cellular infiltrate was observed in the ND-silk treated wounds, compared to saline controls (p=0.07, Figure 6(i)). We observe this as an expected physiological response to the application of a nanoparticle such as NDs, an external agent that is likely to cause the infiltration of immune cells. However, this response is minimal with ND-silk and, overall, does not negatively affect wound healing.

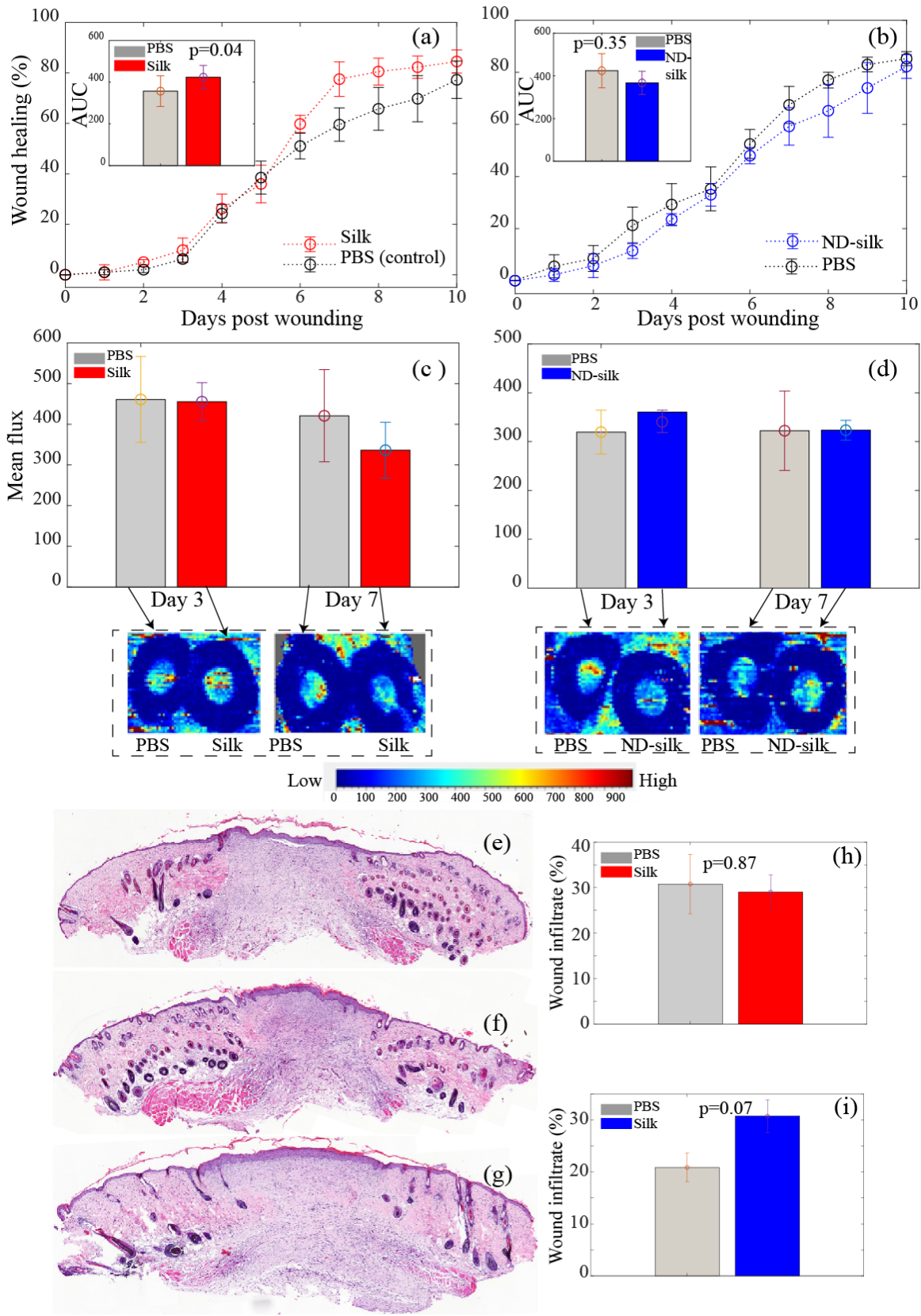

**Figure 6:** Compatibility testing of silk and ND-silk hybrid platforms in a murine wound healing model. Wound closure from day 0-10 of post wounding as a percentage of the wound on day 0 in (a) PBS vs silk membranes and (b) PBS vs ND-silk hybrid membranes. The bar graph insets represent means of the area under curve (AUC) for each wound in each mouse. The p values were derived from a paired t test for a and b. The wound blood flow perfusion measured using laser Doppler imaging on day 3 and day 7 of post-wounding in mice with (c) PBS vs silk and (d) PBS vs ND-silk hybrid membranes. The representative images from perfusion studies for day 3 and 7 for PBS vs silk treated wounds and PBS vs ND-silk covered wounds are shown inside dashed boxes. The scale bar indicates the range of perfusion from low (dark blue) to high (red) blood flow. Representative histological cross sections from the wounds stained with Haematoxylin and Eosin for (e) PBS, (f) silk membrane and (g) ND-silk membrane treated wounds, on day 10 post-wounding. The wound cellular infiltrate calculated for each wound as a percentage of the total wound granulation area; (h) PBS vs silk membranes

and (i) PBS vs ND-silk membranes. The *p* values were derived from an unpaired t-test for (h) and (i). All data are presented as mean ± SEM (standard error of the mean) with n=3-4 mice.

## 2.6 Antibacterial activity

To assess the antibacterial properties of the silk and ND-silk hybrid scaffolds, we fabricated electrospun silk fibroin fibres containing low (0.25 mg/mL) and high (0.5 mg/mL) concentrations of NDs. The two ND-silk membranes are denoted as ND$_{0.25}$-silk and ND$_{0.5}$-silk, respectively. The results are shown in Figure 8.

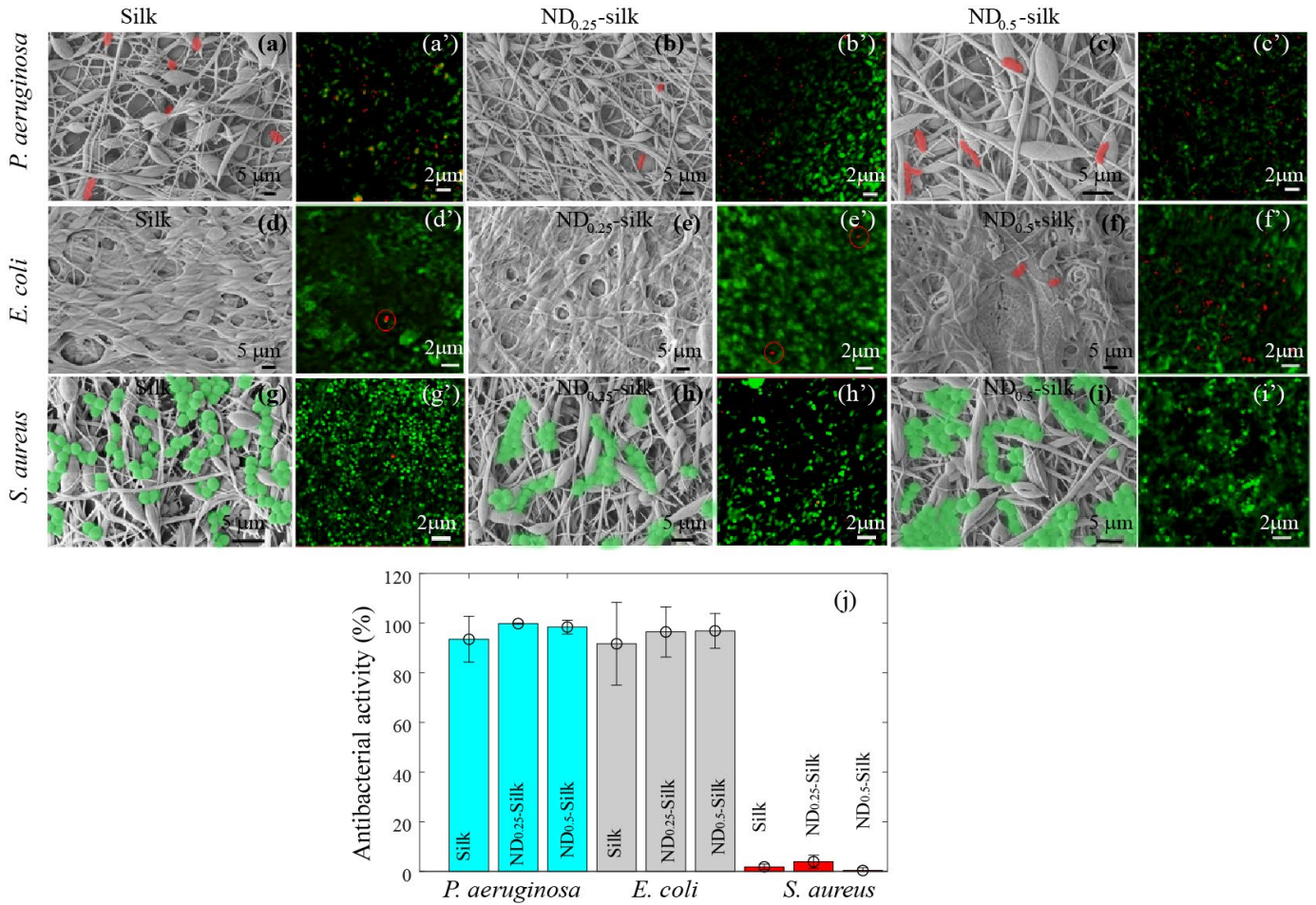

**Figure 7:** Antibacterial activity of ND-silk membranes. *P. aeruginosa*, *E. coli* and *S. aureus* cell morphology on the (a, d, g) silk only, (b, e, h) ND$_{0.25}$-silk and (c, f, i) ND$_{0.5}$-silk membranes. The bacteria are highlighted by red (non-viable) and green (viable) colours. A very low number (red) of *P. aeruginosa* and *E. coli* can be seen on the membranes in (a)-(f). In contrast a large numer (green) of viable *S. aureus* bacteria were observed to adhere and proliferate on the silk-ND fibrous membranes. Confocal laser scanning (CLSM) microscope images of *P. aeruginosa*, attached onto (a') Silk only, (b') ND$_{0.25}$-silk and (c') ND$_{0.50}$-silk. CLSM images of *E. coli* incubated with (d') Silk only, (e') ND$_{0.25}$-silk and (f') ND$_{0.50}$-silk. CLSM images of *S. aureus* bacteria incubated with (g') silk only, (h') ND$_{0.25}$-silk and (i') ND$_{0.50}$-silk. Fluorescing round green objects indicate viable bacteria and fluorescing red colour indicates. (j) Antibacterial activity of silk-ND samples. High levels of *P. aeruginosa* and *E. coli* cell death were observed across all samples. No significant differences were observed between the antibacterial activity of silk only and ND-silk hybrid membranes towards both Gram-negative bacterial strains. Very low levels of cell death were observed for the *S. aureus* incubated with silk and ND-silk membranes.

The antibacterial properties of the silk membranes with and without NDs were tested against three pathogenic bacteria of clinical relevance, *Pseudomonas aeruginosa*, *Escherichia coli* and *Staphylococcus aureus*. These bacteria represent the two major prokaryotic taxa, Gram-negative and Gram-positive, respectively, and are major players in skin wound infections[37]. Silk, ND$_{0.25}$-silk and ND$_{0.5}$-silk membranes were incubated with suspensions of bacteria for 24 h, after which, the membranes were washed to remove non-adherent bacteria. Bacterial cell attachment and viability were assessed using SEM and confocal laser scanning microscopy (CLSM), respectively.

Analysis of the SEM micrographs in Figure 8 confirm that *P. aeruginosa* and *E. coli* were unable to attach to the membranes in great numbers, and their cellular morphology appeared flattened and damaged (Figure 8(j)-(k) and indicated by the bacteria false-coloured red in SEM micrographs). High magnification SEM micrographs of *P. aeruginosa* and *E. coli* on ND-silk membranes are shown in Figure S9 of the Supporting Information, to show their damaged cell morphology. The attachment of both Gram-negative strains of bacteria to the silk membranes was less than 1000 cells/mm$^2$ quantified per sample. In particular, the attachment of *E. coli* to silk and ND-silk membranes was below detection limit (less than 200 cells/mm$^2$).

In contrast, *S. aureus* cells appeared to exhibit a healthy morphology and even displayed 3-dimensional growth (Figure 8(g)-(i), cells false-coloured in green). The attachment propensity of *S. aureus* bacterial cells to the membranes was much higher than *P. aeruginosa* and *E. coli* bacterial cells, as shown in SEM images of Figure 8(g)-(i). Interestingly, *S. aureus* attachment decreased as the proportion of NDs in the silk membrane samples was increased. This may also be due to the surface topography change. *S. aureus* are known to require a certain degree of surface roughness in order to successfully colonise a surface [38].

For CLSM, the fluorescence labelling of bacteria with propidium iodide and Syto 9® revealed non-viable bacteria by a red fluorescent signal and healthy, viable bacteria by a green fluorescent signal as shown in Figure 8(a')-(i'). Silk fibres are also auto-fluorescent in green when excited with the 488 nm laser. CLSM imaging confirmed that all three silk, ND$_{0.25}$-silk and ND$_{0.5}$-silk membranes were highly biocidal towards both *P. aeruginosa* and *E. coli* cells (Figure 8(j)). Statistically, very small differences in antibacterial activity were determined across silk-only and silk-ND hybrid materials with increasing concentrations of ND content as shown in Figure 8(a)-(f) left to right. The ND-silk membranes achieved an average bactericidal efficiency of ~99% against *P. aeruginosa* and ~95% against *E. coli*. The incredibly high rates of bacterial death hence may be attributed to the surface PEO content of the silk membranes, as discussed in Section 2.1, which is reported to be selectively antibacterial towards Gram-negative bacteria [39-40].

In sharp contrast to the extremely high antibacterial efficiency towards Gram-negative bacteria, the silk and ND-silk hybrid membranes did not display antibacterial activity towards Gram-positive *S. aureus* bacteria, as exhibited in Figure 8(g)-(i). Almost all attaching cells were viable with only 1-4 % of bacteria identified as non-viable on any of the silk membrane samples. These findings are consistent with previous studies that showed Gram-positive bacteria, such as *S. aureus*, survive significantly better in the presence of PEO solutions than Gram-negative bacteria [41].

Although *Staphylococcus* (Gram-Positive bacteria) are responsible for 40-60 % of skin wound infections, however, *P. aeruginosa* are alarmingly causing multi-drug resistant infections specially in hospital settings[42]. Gram-negative bacteria, such as *P. aeruginosa* cause ~20% of skin infections. In an additional 20% cases, *P. aeruginosa* can appear as a pathogen in a mixed microbial infection. Moreover these bacteria are much harder to treat with antibiotics, because of the low permeability of its cell wall and its ability to acquire and express multiple resistance mechanisms [43].

## 3. Conclusions

In conclusion, the current work reports hybrid ND-silk membranes as multifunctional wound dressings capable of *in situ* temperature monitoring (as a vital wound biomarker), healthy cell growth and selective antibacterial resistance while providing wound closure and healing. Electrospinning yielded highly porous hybrid ND embedded silk membranes. The NDs were able to fluoresce brightly inside silk fibres and added thermal stability to the silk. The *in vitro* test revealed that the surface of the hybrid membranes enables eukaryotic cells attachment and promote the growth of healthy cell. Inside silk fibres, the NV$^-$ centre in NDs provided a greater shift in the zero field splitting 120±20 kHz/K as compared to NDs. This increased frequency shift will lead to enhanced temperature resolution and sensitivity for biological measurements. When applied to an *in vivo* model of wound healing, the ND-silk hybrid membranes enabled wound healing and the presence of NDs as foreign agents did not cause any adverse effects on wound healing and closure. Moreover, the addition of NDs elongated the biodegradation of silk by 3-4 days. Both silk only and ND-silk membranes were found to be highly biocidal towards major skin wound infecting bacteria, *P. aeruginosa* and *E. coli*, (~91 % non-viable cells on silk-only membranes and ~97% non-viable cells on ND-silk membranes). However, the antibacterial efficacy towards *S. aureus* bacteria was negligible. In conclusion, hybrid ND-silk can be used as a multifunctional platform for biosensing as well as provide a network of fibres that supports a healthy wound healing process.

A variety of other sensing and therapeutic agents can be embedded in the membranes for future research for a therapeutic closed loop system with the potential to provide patients with an improved wound treatment alternative. These smart membranes have a number of advantages over the visual assessment approach that is currently used in clinical practice. Firstly, continual measurement of wound temperature is an accurate way to monitor wound quality, compared to visual assessments that can be highly subjective. Secondly, smart membranes also allow for the detection of early signs of wound inflammation and infection. The advantage of early detection is that wound therapies will be more effective. Furthermore, this can be done without the need to remove the dressing and will continually inform wound care specialists about the need to change dressings earlier or later, more frequently or less frequently. Further work and progresses in fibre optics will allow a successful ND-silk wound monitoring device that will have strong potential in chronic wound diagnosis and treatment.

## 4. Experimental

*4.1 Materials*: Hydroxylated 100 nm NDs irradiated with $NV^-$ centre were obtained from ADAMAS as received.

*4.2 Preparation of PEO blend silk fibres: B. mori* silk fibroin solution was prepared according to the reported method [4]. Silk/PEO blends in water were prepared. A 5 w/v% PEO (900,000 g/mol) solution was mixed with 6.5 w/v% silk solution and 0.05-0.1 w/v% NDs. The weight per volume ratio for dry silk fibroin to PEO in the mixture was 6.5:1. To compare the properties of ND-silk hybrid membranes with the silk only samples, we produced two different mixtures by adding 300 μL (13% of the total weight) of PEO in each of (i) 1.5 mL (87% of the total weight) silk with 0.5 mg/mL concentration of NDs and (ii) 1.5 mL silk only. The blended solution was slowly stirred overnight at 4 °C to obtain a homogenous solution. The solutions are then subjected to electrospinning for the formation of silk only and ND-silk hybrid membranes.

*4.3 Electrospinning*: The silk and PEO blends with or without NDs were transferred to 5 ml syringes. The syringe was connected to an 18-gauge needle and mounted on a motor-controlled syringe pump inside a protected chamber at 25 °C. A high positive voltage was applied to the needle through an alligator clip and a collection surface was placed at 12 cm from the tip of the needle. A conductive aluminium collection plate was used which was electrically grounded. Each of the three mixtures silk, $ND_{0.25}$-silk and $ND_{0.5}$-silk were pumped out of the needle with a flow rate of 0.03 mL/min. An electric potential of 12 kV was applied between the needle and the collection plate to obtain a stable jet. The charged jet travels towards the collection plate and deposits in the form of fibrous membranes. The fibres were collected until a desired thickness of ~ 300 μm was achieved. The membranes deposited on glass cover slips were then left in a 90 v/v% methanol/water solvent for 10 min to induce crystallization of β-sheets, resulting in water insoluble membranes. The membranes were finally washed with Milli-Q water and dried in a fume hood for further characterization.

*4.4 Confocal microscopy and ODMR:* The presence of NDs inside the cell-cultured silk-ND membranes was optically examined via a custom-built scanning fluorescence confocal microscope. A 100× air objective (NA = 0.9) was used to deliver continuous-wave 532 nm laser (Laser Quantum, gem 532) to illuminate the sample. Fluorescence generated from silk embedded $NV^-$ centres was collected through a 532 nm dichroic mirror and a 532 nm notch filter to separate the NV fluorescence and the excitation laser, and then coupled into a multimode fibre beam splitter for signal readouts. The collected fluorescence signal was detected by avalanche photo diode (APD, SPCM -AQRH -14) for imaging and temperature sensing. The spectral analysis was implemented through a commercial spectrometer (Princeton Instruments, SpectraPro with a PIXIS CCD camera) [44]. MW radiation for spin manipulation was generated using a MW generator (Rohde & Schwarz) and amplifier and send to an omega-shape resonator. A Peltier stage was used to induce temperature changes in the hybrid membrane within the window of 20-50 °C.

*4.5 Cell culture & in-vitro cell viability:* Human skin keratinocyte (HaCaT) cell line was used. Cells were maintained in Dulbecco's Modified Eagle Media (DMEM), supplemented with 10% foetal bovine serum and 1% penicillin-streptomycin. Cells were grown at 37 °C, in a humidified atmosphere of 95% relative humidity and 5% $CO_2$. For cell viability studies, 1 x $10^5$ cells/well were seeded on to the electrospun membranes on glass cover slips contained in 6 well plates. Bare cover slips, without any silk were used as controls. Cells were allowed to grow for 48 h and then treated with a final concentration of 0.05mg/mL of resazurin solution in dye-free DMEM and incubated at 37 °C for 15 minutes. The fluorescence was then measured on a microplate reader by exciting at a wavelength of 560 nm and collecting the emission at 590 nm. For confocal imaging, the cells were grown at the same seeding density, followed by fixing in 4% paraformaldehyde solution in phosphate buffered saline for 10 minutes.

*4.6 SEM:* The Thermo Scientific™ *Verios* G4 SEM was used at an accelerating voltage of 3 kV to perform the structural analysis of the membranes. Prior to scanning, the samples were sputter coated with ~5 nm Pt layer.

*4.7 Differential scanning calorimetry (DSC)*: DSC was performed on about 5 mg of material in an open crucible using a TA2920 calorimeter (TA Instrument) under flushing nitrogen (100 ml/min) in the range 20–250 °C. Heating the samples for DSC from 20.00°C to 250.00°C was performed at 10.00°C/min

*4.8 In vivo implantation in wound:* All animal care and handling procedures were performed in accordance with the Australian Code for the Care and Use of Animals for Scientific Purposes (2013), as approved by the Animal Ethics Committee of the South Australian Health and Medical Research Institute (SAHMRI), Australia. The C57BL6 mice were bred in-house at the SAHMRI Bio-resources animal facility, fed on a chow diet and water provided *ad libitum*. The biocompatibility of the silk and silk-nanodiamond scaffolds on wound healing were tested on 8-10-week-old mice, using a previously described murine model of wound healing [34]. In summary, the operative region was prepared under general anaesthesia (with 5% isoflurane) by shaving the skin between the two shoulder blades up to the base of the neck, cleaned with alcohol and povidone iodine. Two circular full thickness wounds were created on either side of the midline using a sterile biopsy punch (6 mm in diameter) and removing the skin. The right wound was covered with a circular (4 mm in diameter) silk or silk-nanodiamond scaffold with 10 μL Phosphate buffered saline (PBS). The left wound was covered with 10 μL of phosphate buffered saline (PBS) as the vehicle control. A silicone splint, supported by adhesive and nylon sutures, was applied to both wounds. Both wounds were covered with a transparent occlusive dressing (Opsite™). The mice were monitored daily for weight loss and pain relief was provided with subcutaneous buprenorphine (0.1mg/kg/day). Mice were followed for 10 days post-wounding and the wound area was calculated daily using the mean diameter from X, Y and Z axes measurements. Blood perfusion in each wound was determined using laser Doppler imaging (moorLDI2-IR, Moor Instruments, Devon, UK) on day 3 and 7 post-wounding. On day 10 post-wounding, the mice were humanely killed using isoflurane overdose. Wounds were excised and processed for histological analysis as described previously [34]. Haematoxylin and Eosin (H&E) stain was used to identify the wound infiltrate in wound cross-sections (5μm). The inflammatory wound cellular infiltrate was calculated as the percentage of positively stained cell nuclei in the granulation tissue area using Image-Pro Premier 9.2 (Media Cybernetics, MD, USA) software.

*4.9 Antibacterial properties:* Bacterial suspensions of *Pseudomonas aeruginosa* ATCC 9721, *Escherichia coli* ATCC 11775 and *Staphylococcus aureus* CIP 68.5$^T$ with an optical density ($OD_{600}$) of 0.1 in nutrient broth (Oxoid) were made up by diluting 1 loopful of freshly subcultured bacteria (grown overnight on nutrient agar from stock at 37 °C). Silk and silk-ND membrane samples were cut to size (~1×1 $cm^2$) and incubated in 1 mL (enough to submerge the samples) of bacterial suspension in sterile 24-well plates for 18 h at 25 °C in dark and static conditions. Samples were then washed gently with distilled water, stained with LIVE/DEAD Baclight containing a mixture of Syto 9® and propidium iodide for 30 min and viewed under confocal laser scanning microscopy (CLSM) using an Olympus FV3000 instrument. Syto 9® and propidium iodide are membrane permeable and non-membrane permeable, respectively, nucleic acid stains that differentiate between viable (green) and non-viable (red) cells. Counts of live and dead cells were quantified using Matlab software, CellC.

SEM was performed to assess bacterial cell attachment and morphology on silk and silk-ND membranes. Samples were incubated with bacterial suspensions of *P. aeruginosa*, *E. coli* and *S. aureus*, as above. Samples were washed briefly with MilliQ $H_2O$ and fixed with 2.5 % glutaraldehyde for 30 min before dehydrating in an ethanol series of 30, 50, 70, 90, 95 and 100% concentration for 10 min each. Finally, the samples were sputtered with gold and imaged using the SEM capabilities of a Raith150 Two machine (Raith, GmbH).

**Acknowledgements**

This work was supported by the Australian Research Council (ARC) through the Centre of Excellence for Nanoscale BioPhotonics (CE140100003) and LIEF grant (LE140100131). The SEM and XPS analysis of samples were performed at the RMIT's microscopic and microanalysis facility (RMMF) with Dr Edwin Mayes and Billy Murdoch. The cell culture work was performed at the RMIT Micro Nano Research Facility (MNRF) in the Victorian Node of the Australian National Fabrication Facility (ANFF). The optical characterization and diamond thermometry measurements were performed within the ARC Centre of Excellence for Nanoscale BioPhotonics Laboratories at RMIT. The in vivo studies were performed at South Australian Health and Medical Research Institute (SAHMRI). Cell culture for section 2.4 was assisted by Suneela Pyreddy. The authors acknowledge productive discussions with Prof Andrew Greentree, Dr Ravi Shukla, Dr Marco Capelli, Dr Matthew Field and Dr Snjezana Tomljenovic-Hanic. A.K and P.R. acknowledges funding through the RMIT Vice-Chancellor's Research Fellowship.

# Supporting Information

# Electrospun nanodiamond-silk fibroin membranes: a multifunctional platform for biosensing and wound healing applications

Asma Khalid[1,2*], Dongbi Bai[1], Amanda N. Abraham[1,2], Amit Jadhav[3], Denver Linklater[1], Alex Matusica[4], Duy Nguyen[1], Billy James Murdoch[1], Nadia Zakhartchouk[1], Chaitali Dekiwadia[1], Philipp Reineck[1,5], David Simpson[6], Achini K. Vidanapathirana[2,5,7], Shadi Houshyar[8], Christina A. Bursill[2,5,7], Elena Ivanova[1], and Brant Gibson[1,2].

1. School of Science, RMIT University, Melbourne, VIC, Australia
2. Australian Research Council (ARC) Centre of Excellence for Nanoscale BioPhotonics (CNBP)
3. School of Fashion and Design, RMIT University, Melbourne, VIC, Australia
4. School of Computer Science, Engineering and Mathematics, Flinders University, Clovelly Park SA, Australia
5. Vascular Research centre, Lifelong Health, South Australian Health and Medical Research Institute (SAHMRI), Adelaide, SA, Australia
6. School of Physics, University of Melbourne, Melbourne, VIC, Australia
7. Adelaide Medical School, Faculty of Health and Medical Sciences, University of Adelaide, Adelaide, SA, Australia
8. School of Engineering, RMIT University, Melbourne, VIC, Australia

*asma.khalid@rmit.edu.au

## Section 1: Electrospinning

The setup used for electrospinning hybrid ND-silk membranes is shown in Figure S1 below.

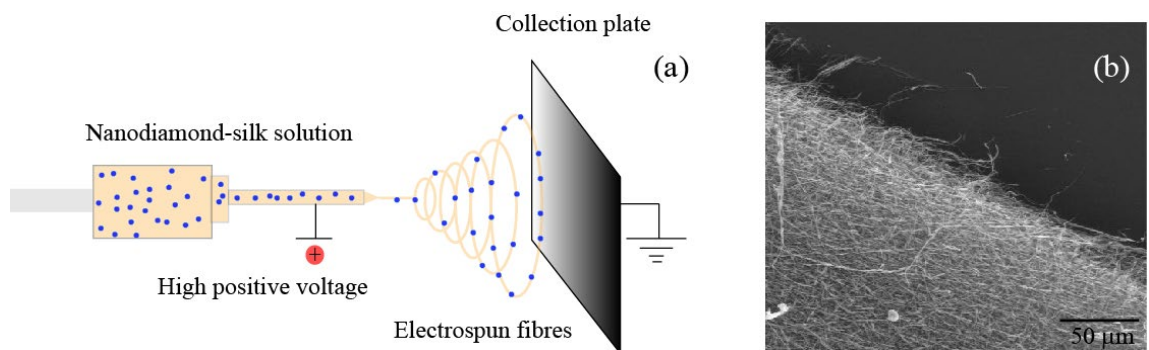

Figure S1: Depiction of the experimental setup. (a) NDs mixed with silk aqueous solution and electrospun as fibrous membranes. (b) A 0.2×0.2 mm$^2$ microscopic image of the collected fibres on the glass coverslip.

## Section 2: X-ray photon spectroscopy (XPS) analysis

An X-ray photon spectroscopy (XPS) analysis of the silk and ND-silk membranes was performed to estimate the surface composition of the membranes before and after methanol treatment step. Figure S1 (a) presents the broad XPS spectrum with C, N, and O forming the surface of silk and ND-silk surfaces before and after methanol treatment. No significant impurities were detected on the surface of the membranes. Figure S1(b) demonstrates the high-resolution $C_{1s}$ profile of the four samples. The binding energies of C–C/C–H (284.8 eV), C–N (285.9 eV), C=O (287.9 eV) are according to reference paper 14 in the main context[1].

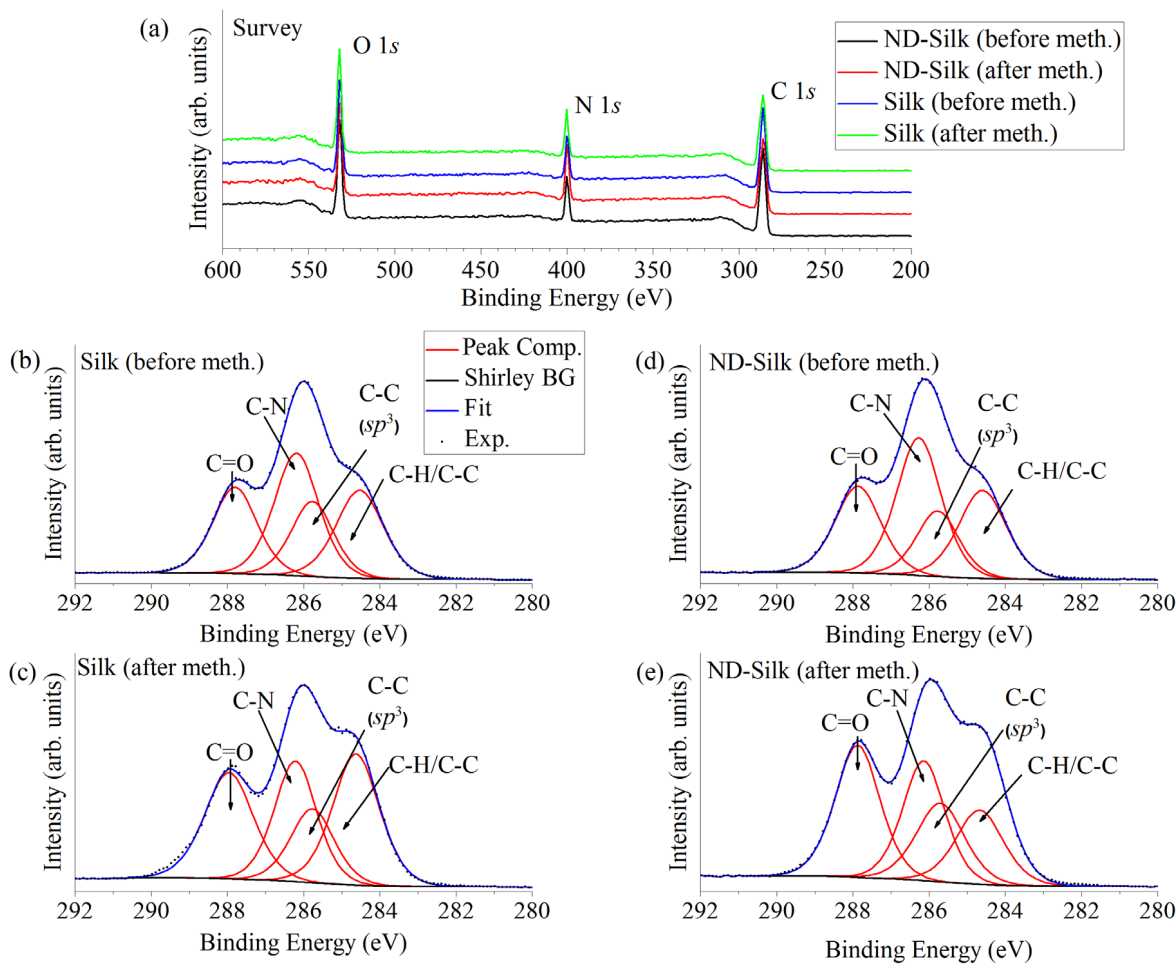

Figure S2: (a) The XPS survey for silk/ND-silk samples before & after methanol treatment. High resolution C1s XPS for (b) Silk before and (c) after methanol treatment, and (d) ND-silk before and (e) after methanol treatment. The C-N bonds (~286 eV) are reduced relative to the C=O (288 eV) & C-H/C-C ($sp^2$) peaks (~285 eV) after methanol treatment.

The results in the above table (Table 1 of the manuscript) also demonstrate that the concentration of each peak is the same with/without NDs to within 3%. It is assumed that the concentration of NDs at the surface is low enough to be negligible for calculation of the PEO concentration. As the remainder of the C-C $sp^3$ content in the samples corresponds to PEO[1], a rough estimate can be obtained by observing the $sp^3$ content in the ND-silk sample pre and post-methanol treatment. An increase of about 17% is hence observed for ND-silk membranes post-methanol treatment for the C-C $sp^3$. This value is in qualitative agreement with the literature[2]. For silk, only membranes, a 22% decrease in silk content (C-N bods) was observed on the surface, as C-N represents the proteins. However, as C-O & C=O bonding is indistinguishable from the C-N peak and the C-H peak is close enough to the $sp^2$ peak to be considered indistinguishable, an accurate measurement cannot be made by XPS.

**Section 3: Cell growth on hybrid membranes**
Human skin keratinocytes (HaCaT) were grown on the surface of the membranes for 48 hours before testing viability using resazurin assay. Bare coverslips with no silk were used as the controls. The SEM results in Figure S3(a) and (b) reveal that the cells attach well to the surface of the membranes with no adverse changes to morphology between cells grown on silk or on ND-silk membranes. Figure S3(c) indicates that there is no significant difference in viability between the control (no sample and silk membrane) and ND-silk hybrid, despite the presence of the NDs.
In both cases, the cells have spread, with filopodia attaching to silk fibres, which suggests that the silk fibres do not hinder cell attachment nor has the presence of NDs inside the silk fibres affected the cell growth.

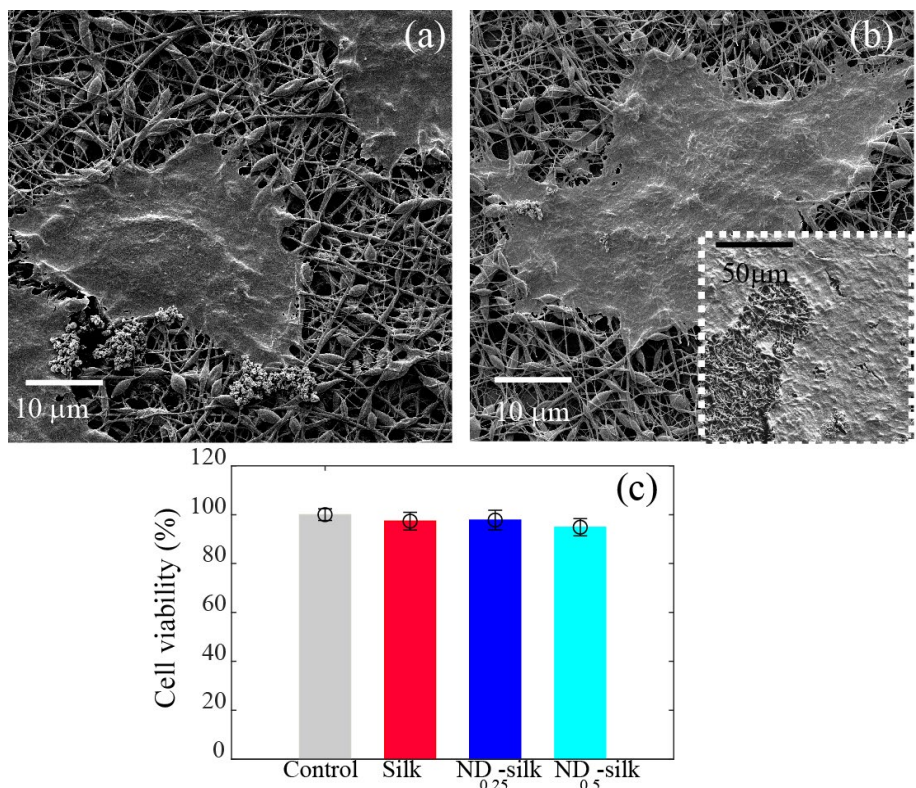

Figure S3: HaCaT cells grown on (a) Silk only (b) ND-silk-diamond (right) fibrous membranes. Scanning electron microscope images of the cells (a) growing on a layer of $ND_{0.5}$-silk fibres. The inset in the white box provides a magnified image of cells. The ND-silk fibrous membrane is approximately 300 μm thick. (c) Cell viability of HaCaT cells grown on the surface of glass coverslip (control), silk, $ND_{0.25}$-silk and $ND_{0.5}$-silk membranes after 48 hours of culture.

Confluency of the cells may be observed from the SEM images after 48 h of cell growth. Within an SEM area of about 200 μm$^2$, we observed a similar density of cells growing on both silk and ND-silk membranes, as shown in the Figure S4 (a) and (b), respectively. This suggests that the cell growth on both these samples were similar to each other. We also provide an overview of a larger area of the sample in Figure S4 (c), where it can be observed that cell confluency is > 90%.

Additionally, cell viability (Figure S3) was determined by resazurin assay. Only viable cells can metabolise the resazurin to the fluorescent resorufin product, which is the analyte measured to determing cell viability [3]. Therefore, it can be inferred that the confluency of viable cells on silk and ND-silk membrane were equivalent to the confluency of cells grown on the control (glass cover slips).

This suggests that silk provides a good matrix for wound dressing applications. While other polymers such as poly-capralactone (PCL) fibres may offer low cytotoxicity[4], they offer low biodegradation rates [5] that limit their long term applications in tissue engineering applications. In contrast silk undergoes enzymatic degradation in the body and the resulting amino acids may be metabolised by the surrounding cells [6].

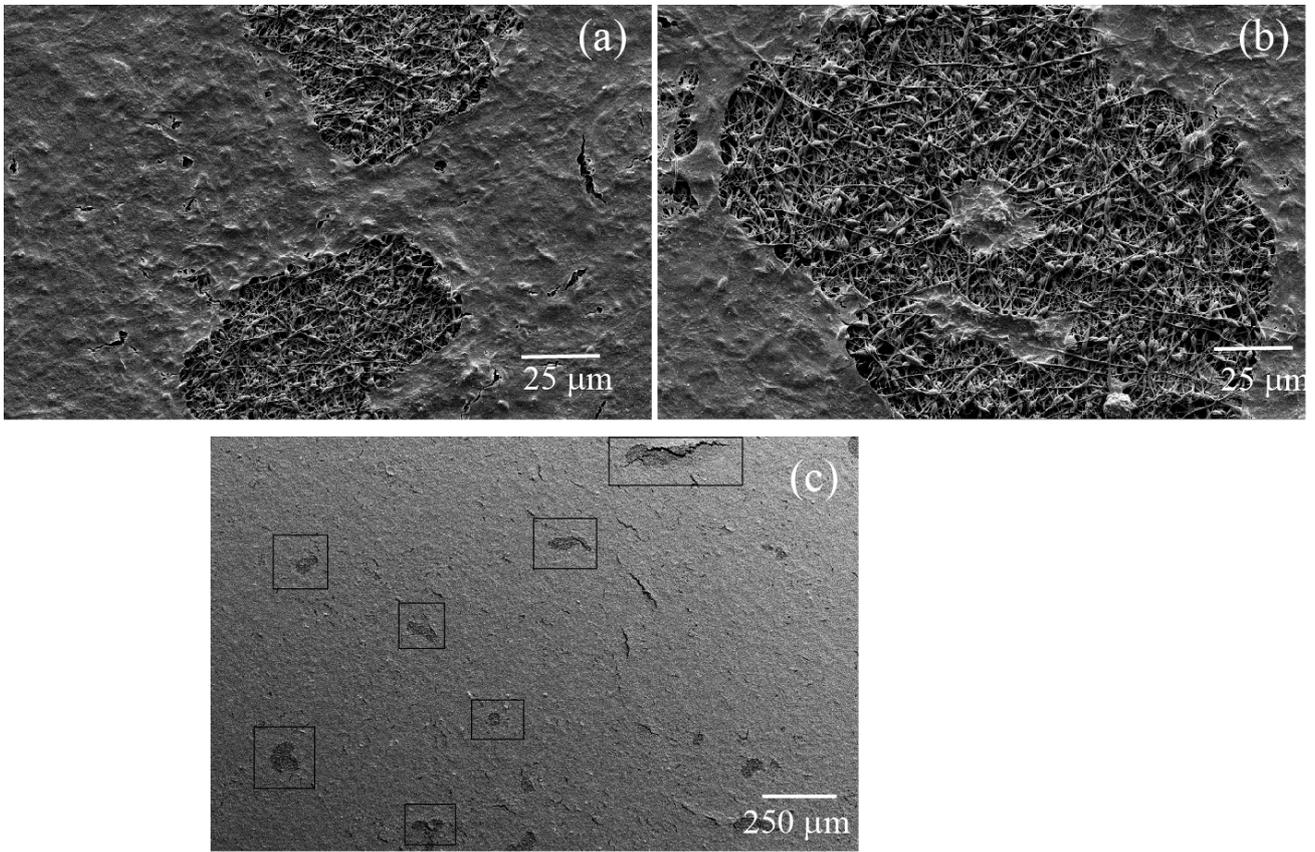

Figure S4: SEM images 200×100 μm² showing a similar confluency for both (a) silk only and (b) ND-silk membranes after 48 h incubation. (c) A larger zoomed out area on the sample where the ND-silk membrane is widely covered by cells except the highlighted areas.

## Section 4: Temperature sensing with ND-silk membranes

To confirm the value of dD/dT parameter for silk embedded $NV^-$ centres, the zero filed splitting $D$ of 6 additional centres was tracked as a function of temperature in a window of 20 °C to ~50 °C. Figure S3(a) and (b) show the calibration curves of the additional six $NV^-$ centres inside ND-silk fibres. The six centres showed resonant frequency shifts of 87 kHz/°C (black circles), 96 kHz/°C (green circles), 113 kHz/°C (magenta dots), 118 kHz/°C (black triangles), 136 kHz/°C (red circles) and 162 kHz/°C (red dots) as shown in Figure S3 (a)-(b). The equations of the straight lines fitted to the experimental data are -0.12T+2872.8 (black), -0.16T+2874 (red) and -0.096T+2871.5 (green) in Figure S3(a). The equations of the straight lines fitted to the experimental data are of Figure S3(b) are -0.09T+2872.5 (black), -0.14T+2873.2 (red) and -0.11T+2872.4 (magenta). The variation in the *dD/dT* parameter could possibly be due to agglomeration of NDs, different number of $NV^-$ centres in each detecting spot and varying z-heights for NDs inside the silk fibres that correspond to variation in the fluorescence of NDs [7-8]. Moreover, different ND emitting spots have a varying height from the heat source, which generates a difference in the temperature they detect.

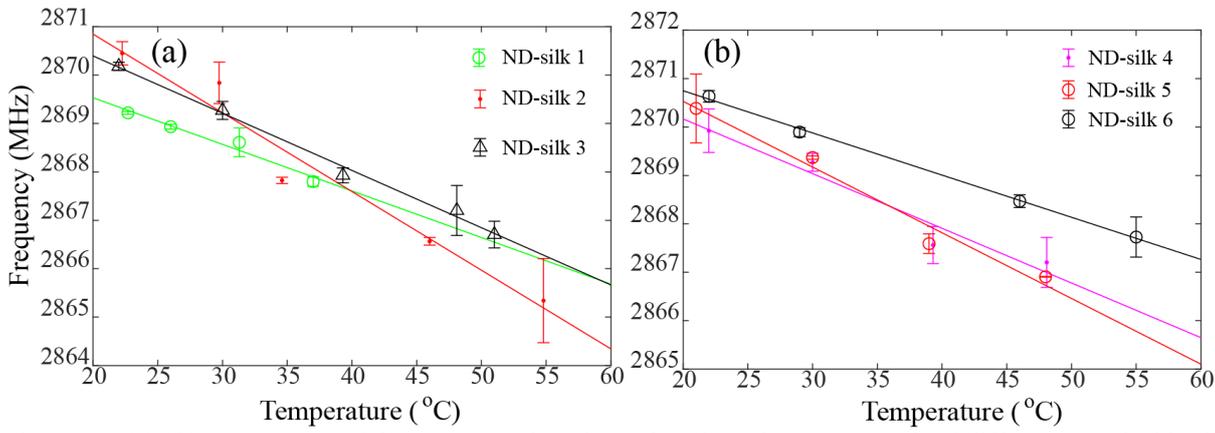

Figure S5: (a), (b) ODMR central frequency *D* plotted as a function of temperature for 6 different NDs inside silk fibres. The data points are represented by asterisk, triangles, circles and dots while lines of the same colour correspond to the linear fit. At each temperature point, 3-5 readings were recorded, and the average central frequency and the error in the mean values are plotted.

**Section 5: Effect of laser power on zero-field splitting frequency**

The effect of increasing laser power on the zero-field splitting *D* was measured, as presented in Figure S4. We experimentally changed the laser excitation power from 1 µW to 40 µW while monitoring the ODMR spectrum from one individual ND emitting spot. The laser power did not induce additional heating effects for the NV$^-$ centres in ND-silk membranes.

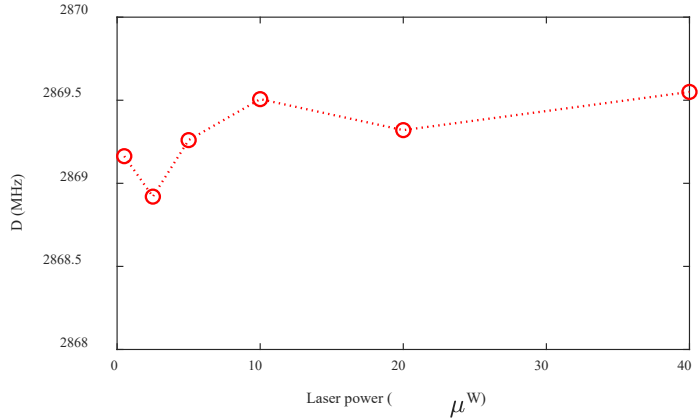

**Figure S6:** Effect of increasing laser power on the value of zero-field parameter D of NV$^-$ centres inside the ND-silk membrane.

**Section 6: Widefield ODMR spectroscopy**

The widefield microscopy was performed on the hybrid ND-silk membranes as shown by the ~100 µm$^2$ image shown in Figure S5(a). The image shows a brightfield (white light) image of the fibrous ND-silk membrane superimposed on its fluorescence image. The silk fibres can be seen as dark grey threads in the brightfield scan while the fluorescence from NDs (excited with 532 nm) is evident with bright red distinct dots in the overlaying fluorescence map. The ODMR data on one of the NDs, Figure S5(b) demonstrate that multisite ODMR can also be performed for temperature monitoring on a biologically active area.

Figure S7(b) shows a widefield ODMR for one of the representatives NDs (shown inside the square in a) at a temperature of 25 °C (blue data) and 37 °C (red data). A zero-field splitting $D_{25}$=2870 MHz was observed at room temperature while a $D_{37}$=2868 MHz shifted frequency was observed at body temperature. Hence the widefield ODMR with these materials is possible and can be explored further for non-invasive temperature sensing.

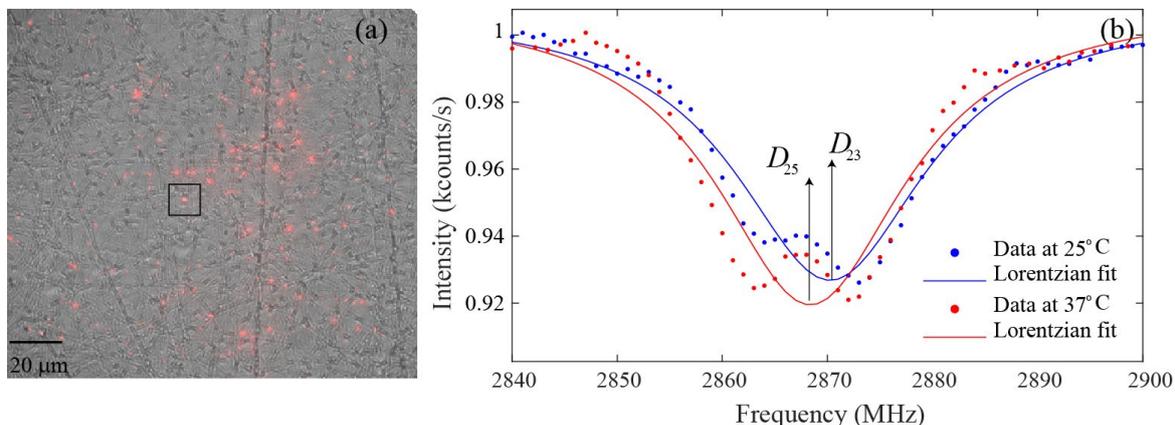

Figure S7: (a) An overlay of Bright field (white light) microscopic and widefield fluorescence images for ND-silk membranes. Fluorescence images were excited with 532 nm laser via a 100× oil objective and a beam expander. (b) ODMR spectra for one of the representative NDs at room and body temperatures, showing a resonant frequency difference of $\varDelta D \sim 160$ kHz/°C.

**Section 7: Murine model of wound healing**
Silk and ND-silk hybrid scaffolds were tested for biocompatibility in an established murine model of wound healing, in comparison to the PBS control. Mice were followed for 10 days post-wounding, and the wound area was calculated daily. The splints and the sutures when replaced when required.

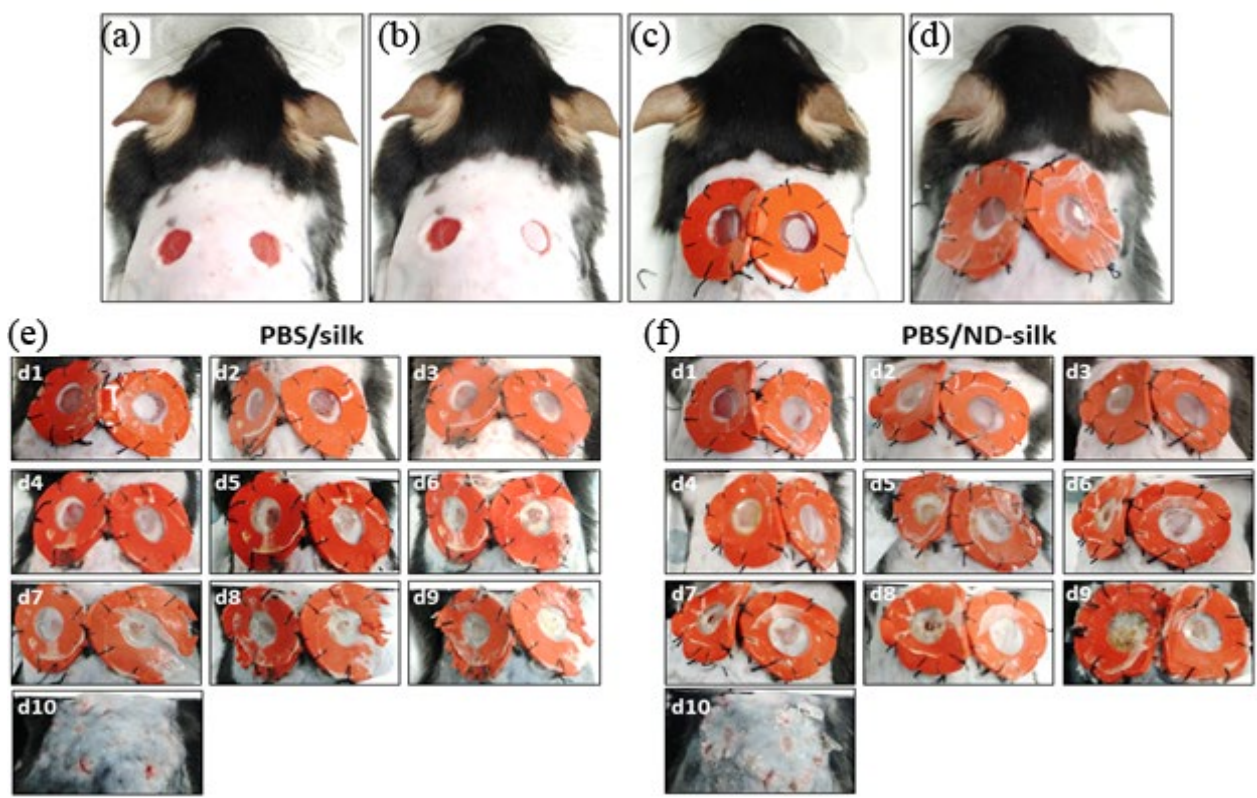

Figure S8: (a) The operative region was prepared under general anaesthesia by shaving the skin between the two shoulder blades up to the base of the neck, and two circular full-thickness wounds were created on either side of the midline using a sterile biopsy punch. (b) The right wound was covered with a circular (4 mm in diameter) silk or ND-silk scaffold with 10 µL Phosphate buffered saline (PBS). The left wound was covered with 10 µL of phosphate buffered saline (PBS) as the vehicle control. (c) A silicone splint, supported by adhesive and nylon sutures, was applied to both wounds. (d) Both wounds were covered with a transparent occlusive dressing (Opsite$^{TM}$). Representative images from the wounds on each day of wound healing (d1-10) (f) from mice treated with PBS and silk and (f) from mice treated with PBS and ND-silk.

**Section 8: Antibacterial properties**
High magnification SEM micrographs of *P. aeruginosa* and *E. coli* on ND-silk membranes are shown in Figure S9, to show their damaged cell morphology.

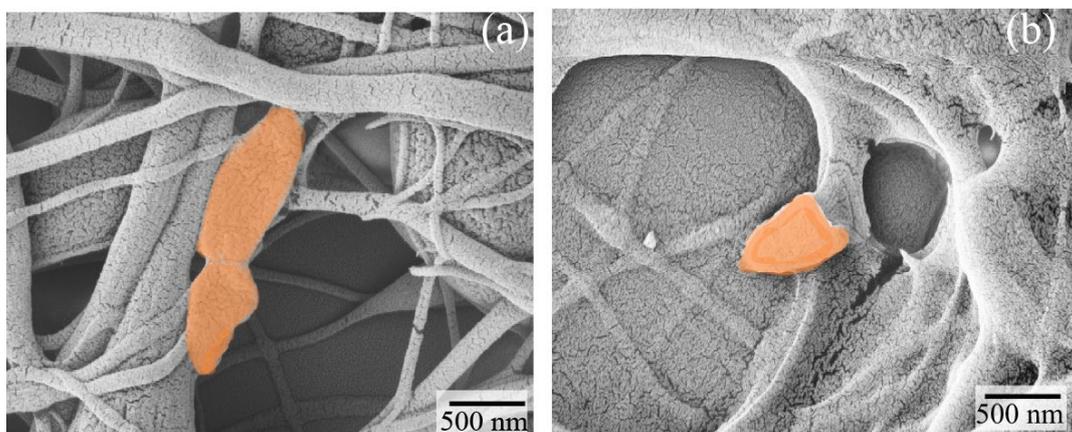

Figure S9: High-magnification SEM micrographs of the compromised cellular morphology for (a) *P. aeruginosa* and (b) *E. coli*. The bacteria are shown in false orange colour.